\documentclass[runningheads]{llncs}
\usepackage[T1]{fontenc}
\usepackage{graphicx}
\usepackage[table,xcdraw]{xcolor}
\usepackage{hyperref}
\usepackage{color, soul}
\usepackage{fontawesome}
\usepackage{booktabs}
\usepackage{pifont}
\usepackage[T1]{fontenc}
\usepackage{circledsteps}
\usepackage{bbding}
\usepackage{marginnote}
\usepackage{todonotes}
\usepackage{subcaption}
\usepackage{enumitem}
\usepackage{float}
\usepackage{tcolorbox}
\tcbuselibrary{breakable}
\usepackage[english]{babel}
\usepackage{microtype}

\definecolor{light-gray}{HTML}{cfd8e8}

\usepackage{multirow}
\usepackage{tikz}

\newcommand{\stpatool}{{SAM}}

\begin{document}
\title{Systems-Theoretic and Data-Driven Security Analysis in ML-enabled Medical Devices}
\titlerunning{Analysis of Security in ML-enabled Medical Devices}
%
\author{Gargi Mitra\inst{1}\orcidID{0000-0001-8011-4590} \and
Mohammadreza Hallajiyan\inst{1}\orcidID{0000-0003-1570-7351} \and
Inji Kim\inst{2}\orcidID{0009-0000-9211-8433} \and
Athish Pranav Dharmalingam\inst{3}\orcidID{0009-0000-7326-4662} \and
Mohammed Elnawawy\inst{1}\orcidID{0000-0002-4367-8060} \and
Shahrear Iqbal\inst{4}\orcidID{0000-0001-7819-5715} \and
Karthik Pattabiraman\inst{1}\orcidID{0000-0003-2380-3415} \and
Homa Alemzadeh\inst{2}\orcidID{0000-0001-5279-842X}}
\authorrunning{G. Mitra et al.}
%
\institute{The University of British Columbia, Vancouver, British Columbia, Canada\\
\email{\{gargi, hallaj, mnawawy, karthikp\}@ece.ubc.ca}\and
University of Virginia, Charlottesville, Virginia, USA\\
\email{\{ddh8jk, ha4d\}@virginia.edu}\and
Indian Institute of Technology Madras, Chennai, Tamil Nadu, India\\
\email{cs21b011@smail.iitm.ac.in}\and
National Research Council Canada\\
\email{shahrear.iqbal@nrc-cnrc.gc.ca}
}

\newcommand{\homa}[1]{\todo[color=green!40]{\textcolor{purple}{\textbf{HA:} #1}}}
\newcommand{\karthik}[1]{\todo[color=pink!40]{\textcolor{blue}{\textbf{KP:} #1}}}
\newcommand{\gargi}[1]{\todo[color=yellow!40]{\textcolor{red}{\textbf{GM: }#1}}}

\maketitle              
\begin{abstract}
The integration of AI/ML into medical devices is rapidly transforming healthcare by enhancing diagnostic and treatment facilities. However, this advancement also introduces serious cybersecurity risks due to the use of complex and often opaque models, extensive interconnectivity, interoperability with third-party peripheral devices, Internet connectivity, and vulnerabilities in the underlying technologies. These factors contribute to a broad attack surface and make threat prevention, detection, and mitigation challenging. Given the highly safety-critical nature of these devices, a cyberattack on these devices can cause the ML models to mispredict, thereby posing significant safety risks to patients. Therefore, ensuring the security of these devices from the time of design is essential. This paper underscores the urgency of addressing the cybersecurity challenges in ML-enabled medical devices at the pre-market phase. We begin by analyzing publicly available data on device recalls and adverse events, and known vulnerabilities, to understand the threat landscape of AI/ML-enabled medical devices and their repercussions on patient safety. Building on this analysis, we introduce a suite of tools and techniques designed by us to assist security analysts in conducting comprehensive premarket risk assessments. Our work aims to empower manufacturers to embed cybersecurity as a core design principle in AI/ML-enabled medical devices, thereby making them safe for patients.

\keywords{AI/ML-enabled medical devices  \and Security assessment \and Safety assessment \and System-theoretic security analysis \and AI/ML security.}
\end{abstract}
\section{Introduction}\label{sec:introduction}

Machine Learning (ML)-driven applications are becoming increasingly popular in the medical field. ML-enabled medical devices (software or software-driven hardware) assist physicians in critical activities such as remote patient monitoring, controlling surgical equipment, automatic drug administration, and preliminary/advanced disease diagnosis~\cite{fdaml}. These tasks require high accuracy and reliability, and the loss of either of these can endanger patient safety. However, the use of ML in interconnected medical devices has expanded the threat surface of medical systems~\cite{chen2020ecgadv,albattah2023detection,lal2021adversarial,9313421,joel2021adversarial,bortsova2021adversarial,menon2021covid,vargas2020understanding,levy2022personalized,chen2022adversarial,yu2023perturbing,nielsen2022investigating,mangaokar2020jekyll,hu2022adversarial,ma2021understanding} making them more vulnerable to cyberattacks. If an adversary compromises such a device, it can force the ML engine to make incorrect predictions or decisions, which can have catastrophic consequences, such as wrong diagnoses and treatments, leading to severe health complications or even the death of the patient.

Detecting and mitigating cyberattacks in ML-based medical applications is significantly more challenging than in traditional systems for two reasons. First, these applications rely on large datasets and often employ complex, unexplainable models, making their behavior difficult to interpret even for developers. Second, they are highly interconnected with third-party devices that collect patient data for real-time predictions, which are subsequently transmitted to downstream systems or patients and physicians for clinical decision-making and treatment. This high degree of connectivity increases the attack surface, while the complexity of ML models complicates attack detection. Adversaries can exploit the vulnerabilities in the ML models and interface devices to poison training data~\cite{mozaffari2014systematic}, inject erroneous inputs during inference~\cite{finlayson2019adversarial}, or modify model parameters through compromised configuration files~\cite{wang2022threats}. We are particularly interested in inference-time false data injection attacks, which are the easiest to execute and most difficult to detect. 
A recent study on the FDA adverse event reports involving ML-enabled medical devices indicated that over 80\% of the reported events were related to data acquisition problems, leading to no data or erroneous data capture~\cite{lyell2023more}. Several studies have also highlighted the vulnerability of data acquisition systems to adversarial examples~\cite{finlayson2018adversarial,han2020deep,bortsova2021adversarial}. 
The safety-critical nature of ML-enabled medical devices makes it crucial to identify and address these security vulnerabilities before deployment.


\begin{figure}[t]
\centering
\includegraphics[scale=0.4]{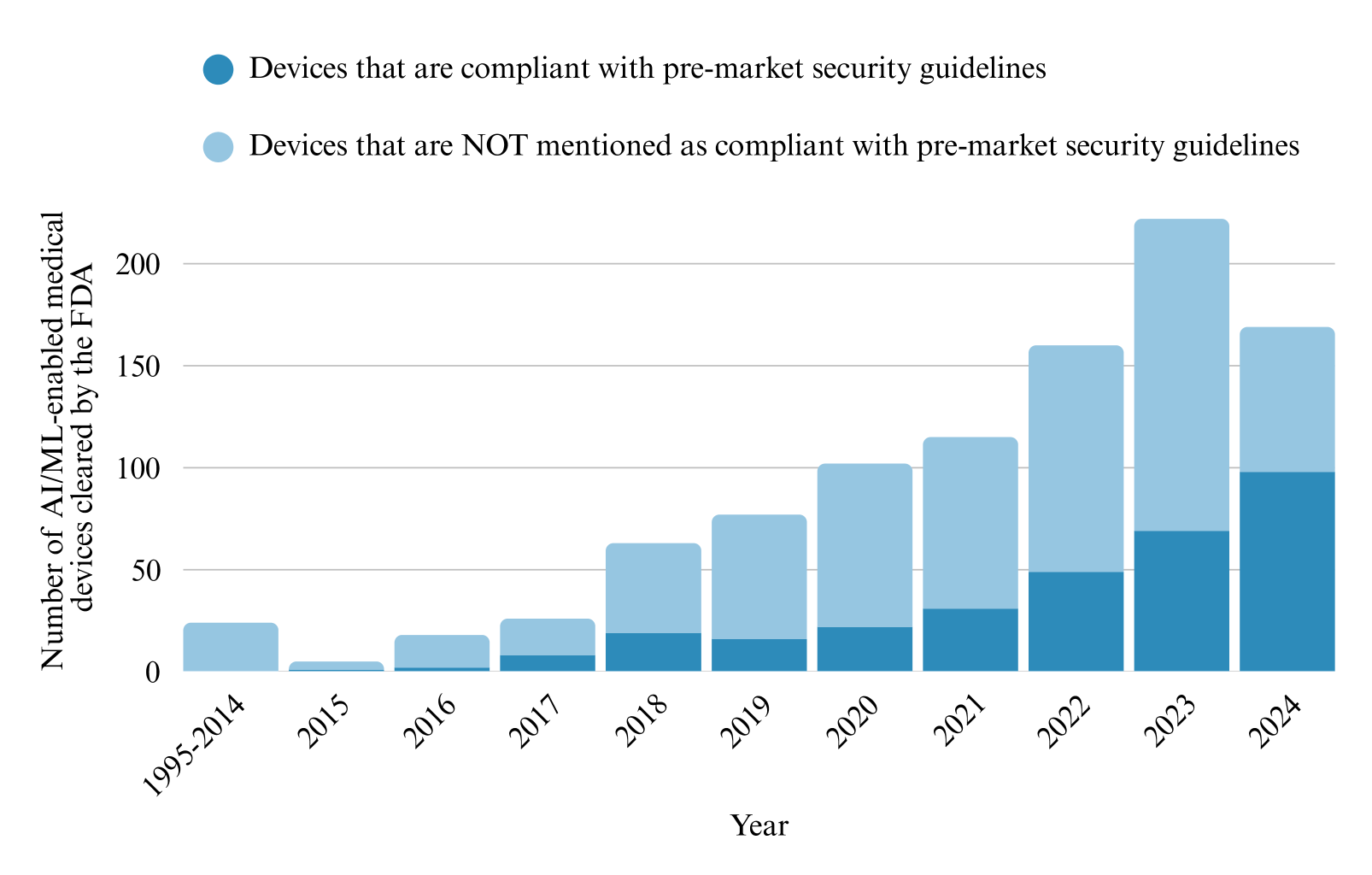}
    \captionsetup{justification=centering}
    \caption{Growing number of AI/ML-enabled medical devices and the rise of security-awareness among device manufacturers (Data as of April 2025)}\label{fig:securitystats}
\end{figure}

In this paper, we focus on pre-market security risk assessment, which is the process of identifying, assessing the severity of, and mitigating potential security risks in a given medical device before it is approved for market release. This process is crucial for ensuring patient safety, regulatory compliance, and cyber resilience, as well as reducing post-deployment threat mitigation costs. 
We inspected the publicly available device summaries~\cite{FDA510k} the manufacturers submitted to the FDA for pre-market approval to determine whether manufacturers conducted security risk assessments for ML-enabled devices.
Our investigation reveals that for over 65\% of these devices, the manufacturers either do not provide any information about the assessment method in their documentation, or employ inadequate assessment methods (see Figure~\ref{fig:securitystats}). In fact, until 2014, no device summary mentioned any security risk assessment.
Among the remaining devices, a few use proprietary mechanisms that make it challenging to assess the adequacy of their approach, while others utilize existing risk assessment techniques. These techniques, as we discuss in the subsequent sections, are insufficient for securing interconnected ML-enabled medical systems. However, on a positive note, there is a growing security awareness among manufacturers, reflected in the increasing mention of security risk assessments in recent pre-market summaries. 

Security practitioners and researchers have made significant efforts in assessing and ensuring the safety and security of medical devices by developing advanced methods for qualitative and quantitative risk assessment (e.g., fault tree analysis (FTA), failure mode and effect analysis (FMEA)) and formal assurance case reports~\cite{Alemzadeh_SP,jetley2006formal,jee2010assurance}) and security analysis~\cite{almohri2017threat}, model-based design and verification~\cite{arney2010toward,pajic2012model,alemzadeh2016}, closed-loop validation~\cite{zhihao2010heartmodel}, encryption, and authentication~\cite{khan2020secure}. However, less attention has been paid to the \textit{end-to-end system security} of ML-enabled medical devices by considering the interactions of the ML-enabled device with other interconnected system components. 
Current security assessment methods primarily focus on algorithm, hardware, software, and firmware vulnerabilities, but they often overlook the \textit{inherent vulnerabilities of the ML models} used in medical devices, how they can be exploited by first exploiting vulnerabilities in interconnected devices, and the potential impact of the ML mispredictions on \textit{patient safety}. \emph{To bridge this gap, it is imperative to perform a holistic system-theoretic analysis of ML-enabled medical systems.} 

In this paper, we first present our experience with developing tools and techniques to automate the extraction of large-scale data on real-world security vulnerabilities and safety incidents for ML-enabled medical devices from public data and knowledge sources. Further, we devise techniques that use this data to enable security practitioners to perform system-theoretic analysis to identify potential threats, new attack paths, and their safety impacts. This will help medical device manufacturers anticipate post-deployment security risks early at design time, assess the severity of the risks, and implement risk prevention and mitigation strategies. For instance, a company developing an ML-enabled device that integrates with third-party commodity off-the-shelf   
cameras can use our techniques to identify known security vulnerabilities in compatible camera models, evaluate the likelihood of their exploitation, and assess potential risks to patient safety based on previously reported failures and adverse events of both ML-enabled and non-ML-enabled devices with the same functionality. Based on these insights, the company can either implement appropriate security measures and safety mechanisms or provide guidance to users to avoid connecting vulnerable cameras to the device. 
We demonstrate our tools and techniques on various ML-enabled medical devices, particularly blood glucose management systems (BGMS), as an example of safety-critical personalized devices with a broad and complex attack surface due to their high levels of connectivity and interoperability.

\section{Background and Motivation}\label{sec:background}
This section provides the technical background required to understand the subsequent sections and the motivation behind our research.

\subsection{AI/ML-enabled Medical Devices} 
As of December 2024, the U.S. Food and Drug Administration (FDA) has authorized more than $1,016$ ML-enabled medical devices across $17$ different medical disciplines (e.g., Cardiology, Ophthalmology, and Gastroenterology)~\cite{fdaml}. These devices can be categorized into two types: Software as a Medical Device (SaMD) and Software in a Medical Device (SiMD). An SaMD is software that can be run on general-purpose computers (e.g., d-Nav for predicting insulin dose for diabetic patients~\cite{dnav}), whereas an SiMD is software that is sold bundled with hardware manufactured by the same company (e.g., GI Genius Intelligent Endoscopy Module~\cite{gigenius}). 
Our analysis of the FDA data shows that while radiological imaging devices are the most common category of FDA-cleared ML-enabled devices (76.5\%), safety-critical devices in clinical chemistry (e.g., BGMS), cardiovascular (e.g., arrhythmia diagnosis devices), and neurology (e.g., surgical procedures planning systems) have relatively higher numbers of reported adverse events (see Figure \ref{fig:MDR_Categories}). Unlike radiological devices, most personalized cardiac monitors and BGMS are mobile-based devices used by patients in the absence of continuous medical supervision. Their compatibility with peripheral devices from multiple brands and various communication protocols creates a broad and complex attack surface, making them highly susceptible to false data injection attacks with potentially severe consequences. These factors make such devices a compelling choice for our evaluation.
Table~\ref{tab:knownmlattacks} shows examples of ML-enabled BGMS, including d-Nav~\cite{dnav}, WellDoc BlueStar~\cite{welldoc}, Dreamed Advisor Pro~\cite{dreamed}, Dario Blood Glucose Monitoring System~\cite{dario}, and the One Drop Blood Glucose~\cite{onedrop} Monitoring System. 


\begin{figure}[t]
    \centering
     \includegraphics[width=\columnwidth,keepaspectratio]{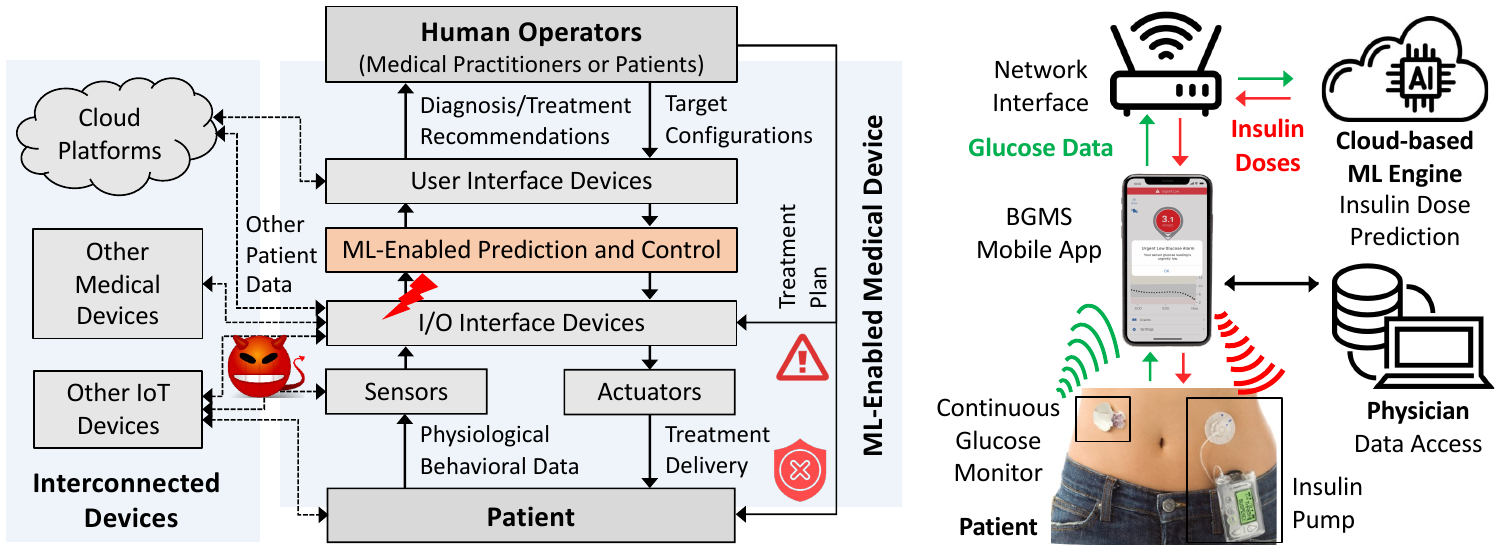}
    \captionsetup{justification=centering}
    \caption{Left: Typical System Control Structure of Interconnected ML-enabled Medical Cyber-Physical Systems. Right: Example ML-enabled Blood Glucose Management System (BGMS) Authorized by the U.S. FDA.} 
    \label{fig:ControlStructure}
    \vspace{-1em}
\end{figure}

\begin{table*}[hbt!]
\captionsetup{justification=centering}
\caption{A study of different FDA-Approved ML-enabled medical devices and their security vulnerabilities that enable false data injection attacks {\footnotesize
$\dagger$: SaMD, $\ddagger$: SiMD, $^*$: Best-guessed ML algorithm, $^\$$YoA: Year of Approval, \CircledTop{L}: Only locally exploitable vulnerability, \CircledTop{R}: Remotely exploitable vulnerability}}
\label{tab:knownmlattacks}
\scriptsize
\setlength\tabcolsep{1pt}
\centering
\begin{tabular}{|c|c|c|c|c|c|c|c|c|c|}
\hline
\rowcolor[HTML]{EFEFEF} 
  \textbf{\begin{tabular}[c]{@{}c@{}}Device \\ Name~\cite{fdaml}\end{tabular}} & \textbf{\begin{tabular}[c]{@{}c@{}}YoA$^\$$\end{tabular}} &
  \textbf{\begin{tabular}[c]{@{}c@{}}Device\\ Function\end{tabular}} &
  \textbf{\begin{tabular}[c]{@{}c@{}}ML\\Technique\\ used\end{tabular}} &
  \textbf{\begin{tabular}[c]{@{}c@{}}Known\\ML\\ attacks\end{tabular}} &
  \textbf{\begin{tabular}[c]{@{}c@{}}Possible third-party\\attack entry points\\\{Known vulnerablity\}\end{tabular}} &
  \textbf{\begin{tabular}[c]{@{}c@{}}Potential\\ impact of\\ mis-\\prediction\end{tabular}} \\ \hline
  \begin{tabular}[c]{@{}c@{}}d-Nav\\System$^{\dagger}$\end{tabular} &
  '19 &
  \begin{tabular}[c]{@{}c@{}}Insulin dose\\ prediction\end{tabular} &
  \begin{tabular}[c]{@{}c@{}}Reinforcement\\learning$^*$\end{tabular} &
  \begin{tabular}[c]{@{}c@{}}~\cite{elshazly2024false}\end{tabular} &
  \begin{tabular}[l]{@{}l@{}}Android\\ vulnerabilities~\cite{CVE-2024-43093}\end{tabular}&
  \begin{tabular}[c]{@{}c@{}}Wrong\\ treatment\\ (Fatal)\end{tabular} \\ \hline  
  \begin{tabular}[c]{@{}c@{}}WellDoc \\ BlueStar$^{\dagger}$\end{tabular} &
   '19 &
  \begin{tabular}[c]{@{}c@{}}Diabetes \\ management\end{tabular} &
  \begin{tabular}[c]{@{}c@{}}Light Gradient\\Boosting  Machine$^*$\end{tabular} &
  \begin{tabular}[c]{@{}c@{}}\cite{amich2021explanation} \end{tabular} &
  \begin{tabular}[l]{@{}l@{}}
  Cloud Service API~\cite{googlecloudcve}\end{tabular}&
  \begin{tabular}[c]{@{}c@{}}Wrong\\ diagnosis\end{tabular} \\ \hline
  \begin{tabular}[c]{@{}c@{}}Dreamed\\Advisor\\Pro$^{\ddagger}$\end{tabular} & 
   '19 &
  \begin{tabular}[c]{@{}c@{}}Diabetes\\management\end{tabular} &
  \begin{tabular}[c]{@{}c@{}}Reinforcement\\learning$^*$\end{tabular} &
  \begin{tabular}[c]{@{}c@{}}\cite{lin2017tactics}\end{tabular}&
  \begin{tabular}[l]{@{}l@{}}Blood glucose\\ meter~\cite{glucosemetercve}\end{tabular}&
  \begin{tabular}[c]{@{}c@{}}Wrong\\treatment\\(Fatal)\end{tabular} \\ \hline
  \begin{tabular}[c]{@{}c@{}}Dario\\BGMS$^{\ddagger}$\end{tabular} & 
  '15 &
  \begin{tabular}[c]{@{}c@{}}Diabetes \\ management\end{tabular} &
  \begin{tabular}[c]{@{}c@{}}k-means\\clustering$^*$ \end{tabular} &
  \begin{tabular}[c]{@{}c@{}}\cite{chhabra2020suspicion}\end{tabular}&
  \begin{tabular}[l]{@{}l@{}}Android\\ vulnerabilities~\cite{CVE-2024-43093}\end{tabular}&
  \begin{tabular}[c]{@{}c@{}}Wrong\\treatment\\(Fatal)\end{tabular} \\ \hline  
  \begin{tabular}[c]{@{}c@{}}One Drop\\BGMS$^{\ddagger}$\end{tabular} & 
   '16 &
  \begin{tabular}[c]{@{}c@{}}Diabetes \\ management\end{tabular} &
  \begin{tabular}[c]{@{}c@{}}Long short-term\\ memory$^*$\end{tabular} &
  \begin{tabular}[c]{@{}c@{}}\cite{sun2018identify}\end{tabular}&
  \begin{tabular}[l]{@{}l@{}}Bluetooth\end{tabular}&
  \begin{tabular}[c]{@{}c@{}}Wrong\\treatment\\(Fatal)\end{tabular} \\ \hline   
  \begin{tabular}[c]{@{}c@{}}Mammo-\\Screen$^{\ddagger}$\end{tabular} & 
   '24 &
  \begin{tabular}[c]{@{}c@{}}Breast cancer\\detection\end{tabular} &
  \begin{tabular}[c]{@{}c@{}}Deep learning\end{tabular} &
  \begin{tabular}[c]{@{}c@{}}\cite{li2023adversarial}\end{tabular}&
  \begin{tabular}[l]{@{}l@{}}PACS server \{\cite{pacscve}\}\CircledTop{R}\end{tabular}&
  \begin{tabular}[c]{@{}c@{}}Wrong\\diagnosis\end{tabular} \\ \hline   
  \begin{tabular}[c]{@{}c@{}}CardioLogs\\ECG\\Analysis\\ Platform$^{\dagger}$\end{tabular} & 
  '17 &
  \begin{tabular}[c]{@{}c@{}}Cardiac \\ arrhythmia\\ detection\end{tabular} &
  \begin{tabular}[c]{@{}c@{}}Deep Neural\\Network (DNN)\end{tabular} &
  \begin{tabular}[c]{@{}c@{}}~\cite{chen2020ecgadv}\end{tabular}&
  \begin{tabular}[l]{@{}l@{}}Portable ECG\\Monitors - \{\cite{ecgvul1}\} \CircledTop{L},\\Cellular network,\\Bluetooth \end{tabular}&
  \begin{tabular}[c]{@{}c@{}}Wrong \\ treatment\\ (Fatal)\end{tabular} \\ \hline
  

GI Genius$^{\ddagger}$ &
  '21 &
  \begin{tabular}[c]{@{}c@{}}Gastro-\\ intestinal\\ lesion\\ detection\end{tabular} &
  \begin{tabular}[c]{@{}c@{}}Convolutional\\ neural\\ networks (CNN)$^*$\end{tabular} &
  \begin{tabular}[c]{@{}c@{}} \cite{hirano2021universal} \end{tabular} &
  \begin{tabular}[l]{@{}l@{}}
  Endoscope\\cameras - \{\cite{endo-1}\} \CircledTop{R}, \\Intranet / Internet \end{tabular}&
  \begin{tabular}[c]{@{}c@{}}Wrong\\ diagnosis\end{tabular} \\ \hline
  

  \begin{tabular}[c]{@{}c@{}}NuVasive\\Pulse\\ System$^{\ddagger}$\end{tabular} &
   '18 &
  \begin{tabular}[c]{@{}c@{}}Neurological\\ monitoring\end{tabular} &
  CNN$^*$ &
  \begin{tabular}[c]{@{}c@{}}~\cite{hirano2021universal} \end{tabular} &
  \begin{tabular}[l]{@{}l@{}}Infra-red sensitive\\cameras - \cite{wang2021can} \CircledTop{L},\\\{\cite{ir1}\} \CircledTop{R}, 
  Internet\end{tabular}&
  \begin{tabular}[c]{@{}c@{}}Mistake\\in\\surgery\\ (Fatal)\end{tabular} \\ \hline
  Air Next$^{\ddagger}$ &
  '20 &
  Spirometer &
  \begin{tabular}[c]{@{}c@{}}CNN$^*$\end{tabular} &
  \begin{tabular}[c]{@{}c@{}}~\cite{hirano2021universal} \end{tabular} &
  \begin{tabular}[c]{@{}c@{}}Bluetooth, Internet\end{tabular}&
  \begin{tabular}[c]{@{}c@{}}Wrong\\ diagnosis\end{tabular} \\ \hline
  \begin{tabular}[c]{@{}c@{}}BrainScope\\TBI$^{\ddagger}$\end{tabular} &
   '19 &
  \begin{tabular}[c]{@{}c@{}}Brain injury\\ assessment\end{tabular} &
  \begin{tabular}[c]{@{}c@{}}Regularized logistic\\regression model\end{tabular} &
  \begin{tabular}[c]{@{}c@{}}~\cite{chen2024adversarial} \end{tabular} &
  Internet&
  \begin{tabular}[c]{@{}c@{}}Wrong\\ treatment\\ (Fatal)\end{tabular} \\ \hline
  \begin{tabular}[c]{@{}c@{}}IDx-DR\\v2.3$^{\dagger}$\end{tabular}&
  '22 &
  \begin{tabular}[c]{@{}c@{}}Diabetic\\ Retinopathy\\ Detection\end{tabular} &
  \begin{tabular}[c]{@{}c@{}}CNN\end{tabular} &
  \begin{tabular}[c]{@{}c@{}}~\cite{hirano2021universal} \end{tabular} &
  \begin{tabular}[l]{@{}l@{}}This device uses the\\Topcon NW200 Fundus\\camera, which comes\\packaged with a PC\\running Windows 7 OS. \\The Windows 7 OS has\\ known vulnerabilities\\- \{\cite{window2}\} \CircledTop{R}, Internet \end{tabular} & 
  \begin{tabular}[c]{@{}c@{}}Wrong \\ diagnosis\\ (loss of\\ vision)\end{tabular} \\ \hline
  \begin{tabular}[c]{@{}c@{}}Iris\\Intelligent\\Retinal\\ Imaging\\System$^{\dagger}$\end{tabular} &
  '15 &
  \begin{tabular}[c]{@{}c@{}}Storage,\\ management\\and display\\of retinal\\images\end{tabular} &
  Deep Learning &
  \begin{tabular}[c]{@{}c@{}}~\cite{yoo2020outcomes} \end{tabular}&
  \begin{tabular}[l]{@{}l@{}}Same as in the case\\of IDx-DR v2.3,\\Internet
  \end{tabular}& 
  \begin{tabular}[c]{@{}c@{}}Wrong \\ diagnosis\\ (loss of\\ vision)\end{tabular} \\ \hline
  \begin{tabular}[c]{@{}c@{}}Paige\\Prostate$^{\dagger}$\end{tabular}&
  '21 &
  \begin{tabular}[c]{@{}c@{}}Cancer\\ diagnosis\end{tabular} &
  \begin{tabular}[c]{@{}c@{}}CNN + Recurrent \\ neural networks\end{tabular} &
  \begin{tabular}[c]{@{}c@{}}\cite{ghaffari2022adversarial}\end{tabular} &
  \begin{tabular}[l]{@{}l@{}}
   Medical scanners\\ 
   - \{\cite{phillips1}\} \CircledTop{L}, Internet
   \end{tabular}& 
  \begin{tabular}[c]{@{}c@{}}Wrong\\diagnosis\\(Fatal)\end{tabular} \\ \hline
  \begin{tabular}[c]{@{}c@{}}Tissue of\\Origin\\ Test Kit$^{\ddagger}$\end{tabular} &
  '18 &
  \begin{tabular}[c]{@{}c@{}}Malignant\\Tumor\\ diagnosis\end{tabular} &
  SVM &
  \begin{tabular}[c]{@{}c@{}}~\cite{ma2021understanding} \end{tabular} &
   Internet&
  \begin{tabular}[c]{@{}c@{}}Wrong\\diagnosis\\ (Fatal)\end{tabular} \\ \hline
\end{tabular}
\setlength\tabcolsep{6pt}
\end{table*}
\textbf{Interconnected medical devices.} The ML models in these devices typically receive inputs from multiple sensory devices that collect various physiological data from a patient's body to predict their condition. Moreover, they can interface with third-party software, cloud platforms, and IoT devices, creating a highly interconnected system. For instance, as shown in Figure~\ref{fig:ControlStructure} (Right), an ML-based diabetes management app such as d-Nav can be installed on a mobile phone. It contains two user-interactive software elements - one for the patient and one for the physician. The system can receive glucose measurement data entered manually into the patient user software or automatically via the cloud from a linked blood glucose meter or continuous glucose monitor (CGM). Some backend components run locally on the phone, while others may be hosted either locally or in the cloud~\cite{dnav}.

\begin{table*}[t!]
\captionsetup{justification=centering}
\caption{Examples of FDA-approved Automated Insulin Delivery (AID) systems that support interoperability in connected components. Modified from~\cite{klonoff2024interoperability}.}
\label{tab:aid-systems-fda}
\scriptsize
\setlength\tabcolsep{2pt}
\centering
\begin{tabular}{|c|c|c|c|c|}
\hline
\rowcolor[HTML]{EFEFEF} 
\textbf{\begin{tabular}[c]{@{}c@{}}AID\\System\end{tabular}} & 
\textbf{\begin{tabular}[c]{@{}c@{}}FDA\\Approval\\Date\end{tabular}} & 
\textbf{Pump} & 
\textbf{\begin{tabular}[c]{@{}c@{}}AGC\\(Control Algorithm)\end{tabular}} & 
\textbf{CGM} \\ \hline
\begin{tabular}[c]{@{}c@{}}Beta Bionics\\iLet Bionic\\Pancreas\end{tabular}& 05/19/2023 & iLet & \begin{tabular}[c]{@{}c@{}}iLet\\Dosing Decision Software\end{tabular}  &\begin{tabular}[c]{@{}c@{}}Dexcom G6,\\Dexcom G7\end{tabular}\\ \hline
\begin{tabular}[c]{@{}c@{}}Insulet\\Omnipod 5\end{tabular}&08/26/2024& \begin{tabular}[c]{@{}c@{}}Omnipod 5\\/ DASH\end{tabular}& SmartAdjust algorithm &\begin{tabular}[c]{@{}c@{}}Dexcom G6,\\Dexcom G7\end{tabular}\\ \hline
Tandem Mobi & 07/11/2023& Mobi & Control-IQ algorithm &\begin{tabular}[c]{@{}c@{}}Dexcom G6,\\Dexcom G7\end{tabular}\\ \hline
\begin{tabular}[c]{@{}c@{}}Tandem t:\\slim X2\end{tabular}& 12/13/2019& t:slim X2 & Control-IQ algorithm &\begin{tabular}[c]{@{}c@{}}Dexcom G6,\\Dexcom G7\end{tabular}\\ \hline
\begin{tabular}[c]{@{}c@{}}Medtronic\\MiniMed 770G\end{tabular}&09/01/2020& MiniMed 770G & SmartGuard technology & \begin{tabular}[c]{@{}c@{}}Guardian Sensor 3,\\FreeStyle Libre 2 Plus\end{tabular} \\ \hline
\begin{tabular}[c]{@{}c@{}}Medtronic\\MiniMed 780G\end{tabular}& 04/21/2023& MiniMed 780G & SmartGuard technology & \begin{tabular}[c]{@{}c@{}}Guardian Sensor 3,\\Guardian Sensor 4\end{tabular}   \\ \hline
\begin{tabular}[c]{@{}c@{}}Twiist\end{tabular}& 04/02/2025&\begin{tabular}[c]{@{}c@{}}Deka\\insulin pump\end{tabular}& Tidepool Loop algorithm & \begin{tabular}[c]{@{}c@{}}FreeStyle Libre 3 Plus \end{tabular} \\ \hline
\end{tabular}
\setlength\tabcolsep{6pt}
\vspace{-1em}
\end{table*}

\textbf{Interoperability in AI/ML-enabled medical devices.}
In recent years, there has been a growing trend toward enhancing \textit{interoperability}, particularly in AI/ML-enabled medical systems. For example, to promote modular integration across BGMS from different manufacturers, the FDA has introduced a framework identifying three essential components in Automated Insulin Delivery (AID) systems, including Alternate Controller Enabled (ACE) pumps, interoperable CGMs (iCGMs), and interoperable glycemic controllers (iAGCs), that can reliably and securely communicate with digitally connected devices to send, receive, and execute drug delivery commands~\cite{FDA_QJI}. Motivated by a broader patient movement towards open and personalized configurations~\cite{tidepool,OpenAPS}, several interoperable AID systems have gained FDA approval. 
Table~\ref{tab:aid-systems-fda} shows seven FDA-approved AID systems among which five incorporate officially designated interoperable components (ACE pump, iCGM, and iAGC). A recently approved iACG, Tidepool \cite{tidepool}, supports a wide range of compatible CGMs and insulin pumps from different manufacturers (such as Medtronic, Tandem, Omnipod, and Dexcom) and is used in a newly FDA-approved AID, called Twiist. Although the current AID devices on the market are not ML-enabled, some of them adopt smart model-predictive control (MPC) algorithms (e.g., SmartAdjust in Omnipod 5~\cite{omnipod}, Control-IQ by Tandem t) that are envisioned to use ML in the near future~\cite{omnipodml}. A similar trend towards growing interoperability in other ML-enabled diabetes management systems is also expected to happen.


This shift towards interconnectivity and interoperability underscores an urgent need for comprehensive system-level security analysis in ML-enabled medical devices. As interconnectivity increases, potential vulnerabilities such as data breaches, insecure interfaces, and compromised control integrity must be proactively addressed through secure-by-design architectures.

\subsection{Security Vulnerabilities in ML-enabled Medical Devices}
The draft guidance containing recommendations for AI-enabled device software functions, published in 2025 by the U.S. Food and Drug Administration (FDA) agency, highlights a number of ML risks that are susceptible to cybersecurity threats~\cite{fda-2025-ai}. These include data poisoning, model inversion/stealing, model evasion, data leakage, overfitting, model bias, and performance drift caused by adversaries. 
The highly interconnected nature of ML-based medical devices provides a multitude of attack vectors to adversaries. 
This is also evident from an increasing number of reported recalls, adverse events~\cite{Alemzadeh_SP}, and security vulnerabilities~\cite{kramer2012security,xu2019analysis,gao2023,health-isac2023} and demonstrated attacks on medical devices across various clinical specialties~\cite{halperin2008pacemakers,li2011hijacking,bonaci2015experimental,alemzadeh2016}. A recent study~\cite{health-isac2023} on over 966 medical devices from 117 vendors found 993 vulnerabilities across medical hardware, operating systems, and software applications. Further, 160 of these had publicly-available exploits that could allow the attackers to target patients and healthcare organizations. 
The majority of these vulnerabilities were found in health IT applications (741) and moderate-risk devices (292) such as medical imaging and monitoring/telemetry devices and infusion pumps. 


\subsection{Threat Model}
In this work, we focus on false data injection attacks, a significant threat to interconnected medical devices. An adversary can force an ML engine to generate incorrect predictions or decisions by injecting carefully crafted malicious data through the data acquisition system during inference~\cite{chen2020ecgadv,ma2021understanding}. 

Preventing such attacks in ML-enabled medical devices is particularly challenging due to their interconnectivity with several other peripheral and sensor devices and networks. Figure~\ref{fig:ControlStructure} (Left) shows the various components of an ML-enabled medical system. 
Adversaries can exploit vulnerabilities in any of the third-party medical and Internet of Things (IoT) devices on the hospital network and/or interface and network devices to find their way into a target ML-enabled device, and inject malicious data into the ML engine even if the ML-enabled device is not compromised. Therefore, it is not enough to secure only the ML-enabled devices. A recent notification by the Federal Bureau of Investigation (FBI) indicated that about 53 percent of connected medical and IoT devices in hospitals have known critical vulnerabilities~\cite{gao2023} that could enable such attacks. In our prior work~\cite{elnawawy2024systematically}, 
we manually analyzed 15 ML-enabled medical devices across various disciplines to examine their ML models, known vulnerabilities, and potential attack vectors in peripheral devices for false data injection during inference. Table~\ref{tab:knownmlattacks} summarizes our findings, revealing that 11 of 15 devices were susceptible to false data injection attacks, with consequences ranging from vision loss to patient death.

\subsection{Systems-Theoretic Safety and Security Analysis}\label{sec:STAMP}Given the highly interconnected nature of ML-enabled medical devices, ensuring their security and safety requires a comprehensive, system-level approach that accounts for complex interactions between components. 
Modern system-theoretic approaches to safety and security of interconnected devices, such as STAMP (Systems-Theoretic Accident Model and Processes)~\cite{leveson2011engineering}, model accidents as complex processes resulting from safety and security constraint violations due to inadequate controls. Systems are represented as \textit{hierarchical control structures}, with each level constraining the one below and communicating their conditions and behavior to the upper levels. System-Theoretic Process Analysis for Security (STPA-Sec)~\cite{young2013systems} and Causal Analysis using System Theory (CAST)~\cite{leveson2011engineering}, built upon STAMP, analyze hardware, software, physical systems, and human operators across control layers to pinpoint threat scenarios, security exploits, unsafe actions, and their causal factors. To assess ML-enabled device vulnerabilities, analysts must (i) model the device's control structure and (ii) identify technologies (e.g., protocols, software, OS, firmware) used in each component. 

While several tools support STPA and STPA-Sec across various domains, we assess their suitability for ML-enabled medical systems based on three key features: (i) applicability to security attacks on ML systems, (ii) applicability to the medical domain, and (iii) automation of causal scenario generation. To this end, we evaluate four state-of-the-art STPA/STPA-Sec tools: \textbf{(1)} A-STPA \cite{abdulkhaleq2014open} and its enhanced version, XSTAMPP \cite{abdulkhaleq2015xstampp} assist in linking unsafe control actions to safety hazards and provide graphical aids for control structure creation but require manual causal scenario identification; \textbf{(2)} SafetyHAT \cite{becker2014transportation} is customized for the transportation sector. This tool offers a graphical interface, data management, and domain-specific guidewords but lacks automated causal scenario identification; \textbf{(3)} WebSTAMP \cite{souza2019webstamp} is a web application designed for STPA and STPA-Sec, that provides structured guidance for identifying hazardous control actions and causal scenarios. It has been applied to Glucose Monitoring and Insulin Pumping System, transportation applications\cite{thomas2011performing}, and chemical reactors\cite{young2017system}; and, \textbf{(4)} SOT \cite{pereira2019stamp} --  this tool helps systems engineers conduct safety and security analyses by leveraging past knowledge to identify causal scenarios. 

In summary, A-STPA, XSTAMPP, and SafetyHAT focus on safety risks from device failures, not malicious attacks. While WebSTAMP and SOT consider security concerns, they still rely on users' knowledge of vulnerabilities and require significant manual effort. Table \ref{tab:related} shows a summary of these tools.

 \begin{table}[t!]
 \captionsetup{justification=centering}
    \caption{Summary of State-of-the-Art STPA/STPA-Sec tools\\ (All these tools generate causal scenarios in semi-automated fashion)}
    \label{tab:related}
 \setlength{\tabcolsep}{1pt}  
    \scriptsize
    \centering
    \begin{tabular}{|m{4cm}|m{3cm}|m{3.5cm}|}
        \hline \centering
        \textbf{Name} & \centering \textbf{Focus} & \centering \textbf{Application Domain} \tabularnewline
        \hline 
         A-STPA, XSTAMPP& Safety & General Purpose \\
         \hline
         SafetyHAT & Safety & Transportation \\
         \hline
         WebSTAMP & Safety/Security & Healthcare, Transportation, Chemical Industry \\
         \hline
         SOT & Safety/Security & Aircraft Systems \\
         \hline
    \end{tabular}
  \setlength{\tabcolsep}{6pt}
\end{table}

Recent papers such as the survey by Qi et al. \cite{qi2023stpa} explore the use of STPA in learning-enabled systems, and introduce DeepSTPA for analyzing ML lifecycle failures, which is beyond our scope. Other recent papers~\cite{qi2023safety,nouri2024welcome,nouri2024engineering} explore the usability of Large Language Models (LLMs) in STPA, highlighting the need for human intervention in generating prompts and validating LLM responses. However, none of these studies focus on medical device security. 

\subsection{Medical Device Databases}
The U.S. Food and Drug Administration (FDA) regulates medical devices sold in the US, and maintains several publicly available databases on premarket and postmarket data about cleared and approved medical devices, including device summary information, approval date, user instructions, and information on Premarket Approvals (PMA), Premarket Notifications (510[k]), Adverse Events, and Recalls. We analyze the following FDA databases to extract the information about medical device technologies and their reported safety and security flaws: 

\textbf{AI/ML-Enabled Medical Devices database}~\cite{fdaml} maintains the information about the FDA-authorized medical devices that incorporate AI/ML across medical disciplines. This data is not comprehensive and only contains releasable information about devices based on information provided in the summary descriptions of their marketing authorization document.

\textbf{Premarket Notifications (510(k)s) database}~\cite{FDA510k} contains the releasable records of premarket notifications submitted by medical device manufacturers for the devices introduced into commercial distribution for the first time or those reintroduced with significant changes. Each record includes device classification and approval information as well as summaries of device functionality and safety and effectiveness information for more recent submissions. 

\textbf{Recalls database}~\cite{FDA_Recalls} contains records of
medical device recalls since November 01, 2002. A recall is a voluntary action that a manufacturer takes to correct or remove from the market any medical device that violates the FDA's laws. Each
record in the database contains the information on a recalled device such as the product name, manufacturer name, number of devices on the market, recall class, FDA determined cause, and the human-written textual descriptions of manufacturer's reason for recall and recovery actions taken to correct the device or remove it from the market. 

\textbf{Manufacturer and User Facility Device Experience database}~\cite{FDA_MAUDE} \textbf{(MAUDE)} is a collection of adverse events of medical devices that volunteers, user facilities, manufacturers, and distributors have reported to the FDA. Each adverse event report contains information such as device and manufacturer names, event type (e.g., Malfunction, Injury,  or Death), event and report dates, and human-written event description and manufacturer narratives, which provide a
short textual description of the incident, as well as any comments made or follow-up actions taken by
the manufacturer to detect and address device problems. 

\subsection{Vulnerability Databases}

We analyze the following publicly available vulnerability databases to identify common threats and security attacks targeting medical devices and peripheral devices:

\textbf{ICS-CERT Alerts dataset}~\cite{ICS-CERT} is developed and maintained by Industrial Control Systems Cyber Emergency Response Team and the United States Computer Emergency Readiness Team (US-CERT). US-CERT is responsible for analyzing and reducing cyber threats, vulnerabilities, disseminating cyber threat warning information, and coordinating incident response activities.

\textbf{MITRE Common Vulnerability Enumeration (CVE) database}~\cite{cvedata}is a publicly accessible registry of known cybersecurity vulnerabilities, maintained by the MITRE corporation. It provides a comprehensive database of vulnerabilities, including those affecting peripheral medical and IoT devices used in medical systems. The data is contributed by software vendors, security researchers, penetration testers, as well as independent researchers.

Despite these publicly available databases, the information on ML-enabled medical devices and the peripheral devices they connect to is only available in a dispersed and unstructured format. It is particularly challenging to (i) extract relevant information on security vulnerabilities and safety impacts from the dispersed data across millions of records in different databases and (2) analyze the free-form natural language text, written by the manufacturers and healthcare professionals, while understanding semantics and the contextual factors involved in the events. 
In the subsequent sections, we describe how some of the tools and techniques we developed alleviate the aforementioned challenges.


\section{Methods}\label{sec:methods}
This section presents our framework for performing holistic system-theoretic analysis of ML-enabled medical systems. The framework comprises a suite of Natural Language Processing (NLP) 
and LLM-aided tools and techniques to assist the systems and control-theoretic security analysis of ML-enabled medical devices. Specifically, we report our experience on the design and validation of tools for semi-automated device modeling and technology identification, information extraction, 
, and systems-theoretic accident causality analysis and attack step generation.

\begin{figure}[t!]
    \centering
    \includegraphics[width=0.98\columnwidth,keepaspectratio]{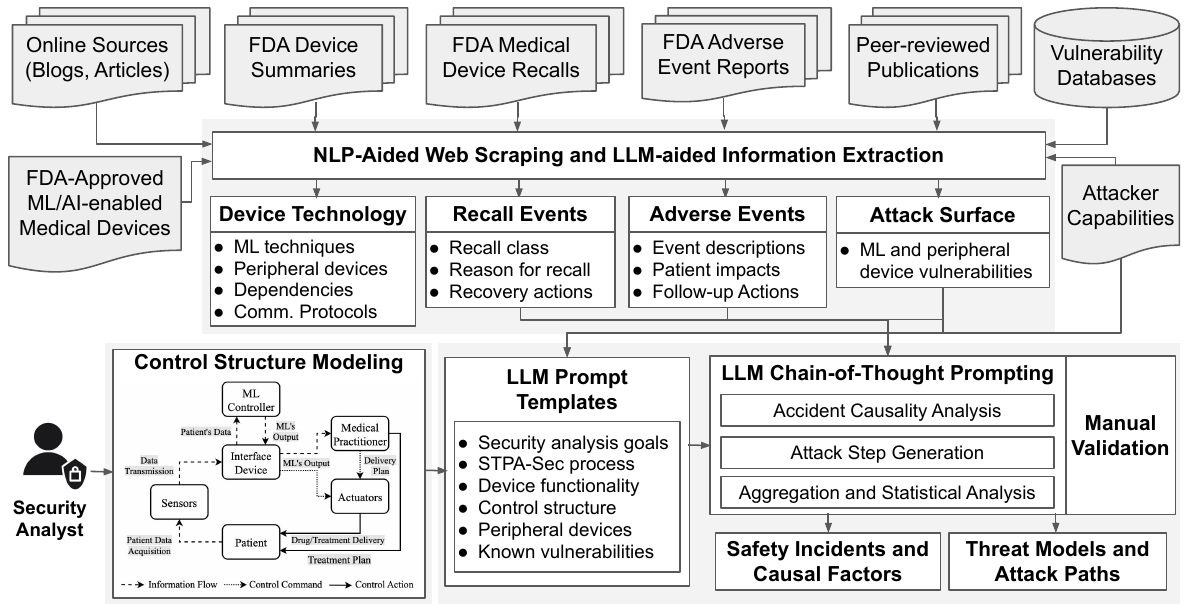}
    \captionsetup{justification=centering}
    \caption{Overall Approach for Systems-Theoretic and Data-Driven Analysis of Safety and Security of ML-enabled Medical Devices}
    \label{fig:Method}
    \vspace{-1em}
\end{figure}

Figure~\ref{fig:Method} shows an overview of our framework, which consists of three main components. 

The \textbf{first} component is a device modeling and technology identification technique. It helps security analysts model the interconnections and communications between the ML-enabled medical device and third-party peripherals as a control structure using a generic control structure template for ML-enabled medical devices. It also assists in identifying all technologies used in connected devices that could serve as potential attack entry points. 

The \textbf{second} component is a set of NLP and LLM-aided web scraping and information extraction techniques to extract and cross-reference the information from publicly available databases. 
Given a natural language description of an ML-enabled medical device, it extracts key details, including the ML technique used, connected peripherals, and device functionality. 
Using this information, it scrapes the web for information on relevant ML vulnerabilities that adversaries could exploit to induce misprediction. Additionally, it interfaces with public FDA medical device and vulnerability databases to identify vulnerabilities in similar medical devices and peripheral devices, as well as recalls and adverse events linked to their malfunctions and safety impacts. 

The \textbf{third} component of this framework is an LLM-based tool that can assist security analysts in systems-theoretic and data-driven safety accident (CAST) and security (STPA-Sec) analysis. This tool integrates the knowledge of CAST and STPA-Sec processes with the extracted information on device technology, control structure, and vulnerabilities and encodes them as customized prompt templates that can guide LLMs in 
generating (i) a comprehensive list of safety issues and causal factors that could lead to patient harm and (ii) attack vectors that adversaries could exploit to deliberately trigger such safety events. 

In the following subsections, we discuss these tools/techniques in detail. We also illustrate how each tool/technique contributes to the overall security assessment, by providing examples of their output when applied to ML-enabled devices
, such as the BGMS in Figure~\ref{fig:ControlStructure}. 

\subsection{Device Modeling and Technology Identification}\label{sec:device-modeling}

To identify all possible attack vectors in a given ML-enabled medical system, a security analyst first needs to model the interconnections of the ML-enabled medical device with the peripheral devices, the data flow between various system components, and understand the technology used by various system components. To enable the systems-theoretic security analysis using STPA-Sec (in Section \ref{sec:stpa-sec}), we adopt the hierarchical system control structures from STAMP (see Section~\ref {sec:STAMP}) for this purpose.

\subsubsection{System Control Structure Modeling.}
\begin{figure}[t!]
\centering
\includegraphics[scale=0.45]{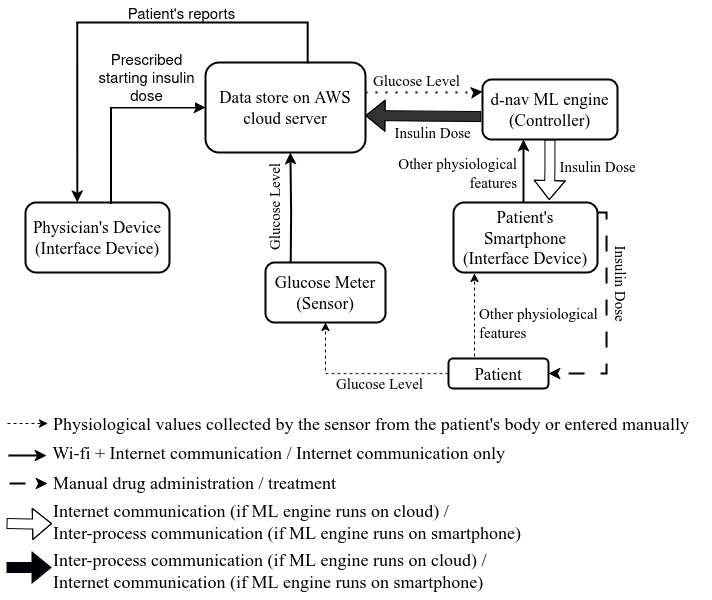}
    \captionsetup{justification=centering}
    \caption{Control Structure of the d-Nav System. Note that, while setting up the system, the ML engine can be configured to run either on the smartphone or on the cloud server.}\label{fig:BGMS}
\end{figure}

In our previous work (SAM ~\cite{Hallajiyan2024SAM}), we partially automated the construction of the system control structure for ML-enabled medical devices by developing a template control structure that contains typical components and interconnections in an ML-enabled medical system. The security analysts could customize this generic control structure by adding/removing necessary components and interactions to match the description of the system under assessment. Once the control structure is built, the security analyst must manually identify the data flows among various system components from the device descriptions. We applied this technique on two ML-enabled medical devices --  (1) d-Nav~\cite{dnav}, a blood glucose monitoring system, and; (2) ABMD~\cite{abmd}, a bone mineral density calculator. We built the control structure and inferred the data flow using the system description provided by the manufacturer, which we obtained from publicly available device summaries submitted to the FDA during the pre-approval process, as well as from information available on the manufacturers’ websites. Figure~\ref{fig:BGMS} shows the control structure for d-Nav~\cite{dnav}, as generated by the technique proposed in SAM. Note that these documents do not follow a standardized format - the information is often dispersed across multiple sites, and the transparency varies across manufacturers. Hence, automating the information retrieval process remains a challenge.

\subsubsection{Technology Identification.}
Once the security analyst builds the control structure, they must identify the technologies used across system components, such as the ML techniques, operating systems, firmware, and communication protocols used by the ML-enabled and connected peripheral devices, to assess the potential security vulnerabilities associated with each of them.

To identify the ML and peripheral device technologies, we have integrated two questionnaires~\cite{sam-questionnaires} into our toolkit~\cite{Hallajiyan2024SAM},  which must be completed by the designers of the ML-enabled device. The first focuses on \textit{compatibility conditions} for each peripheral device in the control structure and needs to be answered by the manufacturer of the ML-enabled device. For example, some blood glucose management systems use Bluetooth to transmit glucose readings from the glucose meter to the glucose management smartphone app, while others require a USB connection for data transfer. Following this, the security analyst must manually identify all commercial peripheral devices that meet the compatibility conditions specified by the ML-enabled device manufacturer. 
The second questionnaire helps analysts identify the \textit{technologies used in the ML-enabled device, and each compatible peripheral device} by covering key technological and operational factors relevant to medical devices. These questions ensure a thorough assessment of potential attack entry points. The factors are categorized into four groups~\cite{Hallajiyan2024SAM}: (i) Human Interaction -- this includes data entry and supervision, data validation, authentication, and anomaly detection; (ii) Communication Protocol -- this includes the exact protocol name, version, and whether it uses encryption; (iii) Electromagnetic Susceptibility -- this includes whether the device is susceptible to electromagnetic radiation, and if so, what its repercussions would be, and if they have any known shielding or mitigation strategy in place; and, (iv) Dependencies on firmware, hardware, OS, and external libraries. This categorization is based on known attack vectors targeting ML-enabled medical devices~\cite{yaqoob2019security}. 

Note that, for a security analyst working for a medical device manufacturing company, obtaining the aforementioned information from the device designers would be straightforward. However, for an analyst working independently or for a third-party company, the manufacturers might be unwilling to provide this information either due to reluctance to spend unnecessary time or effort (as might be the case for peripheral device technologies) or due to confidentiality concerns (as might be the case for ML technique details). In such cases, the analyst can infer compatibility conditions, such as communication links, input devices, and operating systems, from publicly available device descriptions on the FDA website \cite{fdaml} and publicly available information on each peripheral, such as product descriptions on the manufacturer's website. Following this, the analyst could also retrieve a fairly comprehensive list of compatible peripheral devices from third-party information repositories such as TidePool~\cite{tidepool}.
However, identifying the specific ML technologies used is far more challenging, as most manufacturers do not declare them on their website or do not disclose them publicly at all. To assist the analyst under such circumstances, we have developed NLP- and LLM-aided tools as described in the following subsection.

\subsection{NLP and LLM-aided Web Scraping and Information Extraction}\label{sec:infoextraction}
Once the security analyst builds the control structure and identifies the technologies used in the ML-enabled device and its connected components, they must proceed to identify known vulnerabilities in these technologies that might serve as an attack entry point. Today, information about security vulnerabilities, design flaws, and adverse events reported on medical devices is available on the Internet in an unstructured and dispersed manner. This makes it challenging to ensure the coverage of all relevant data during the security assessment process. Our set of NLP-aided web scraping and LLM-aided information extraction tools and techniques assists the system developers and security analysts in extracting and integrating data on all known vulnerabilities and safety issues relevant to the ML-enabled medical device under assessment. This information is also used by our subsequent tools for automated systems-theoretic safety and security analysis. 

\subsubsection{ML Technology and Vulnerability Identification.}
In our latest work~\cite{dharmalingammedaiscout}, we proposed  MedAIScout, a semi-automated NLP- and LLM-aided tool designed to retrieve information on known ML vulnerabilities relevant to ML-enabled medical devices.
MedAIScout works in two steps: \\
(1) \textit{ML technology identification:} Given a description of an ML-enabled medical device, MedAIScout uses NLP techniques to identify key terms related to the device's functionality, ML model type, and data characteristics. Often, the device manufacturers do not publicly disclose the exact ML technique used in their products. In case the security analyst (MedAIScout user) does not have access to a document containing the exact details (such as in the case of third-party analysts), MedAIScout can analyze available information and infer the most likely ML technique by referencing similar devices documented in existing literature. In this work, we sourced the device descriptions from the publicly accessible pre-market device summaries available on the FDA website~\cite{fdaml,FDA510k} and peer-reviewed research articles indexed on Google Scholar.
\\ (2) \textit{ML vulnerability identification:} Next, MedAIScout constructs tailored search queries to retrieve peer-reviewed research articles on attacks targeting the device's ML model. MedAIScout uses local LLMs to differentiate between training-time and inference-time attacks and provides context and explanations for each retrieved article's relevance.

Throughout the device's lifecycle, security analysts can use MedAIScout to track emerging ML vulnerabilities. To the best of our knowledge, it is the first automated tool to retrieve known ML vulnerabilities specifically for medical applications. By applying MedAIScout to five FDA-approved ML-enabled medical devices, we found that MedAIScout successfully uncovered relevant vulnerabilities in four devices, thereby substantially assisting in security analysis. For example, when tested on the One Drop blood glucose monitoring system~\cite{onedrop}, MedAIScout retrieved a peer-reviewed research paper~\cite{tosun2022detection} describing an inference-time attack on a similar system. In this attack, an adversary manipulates blood glucose readings at mealtime by compromising the radio communication between the glucose meter and the controller, leading to incorrect insulin dose recommendations. The paper also proposes an appropriate attack detection technique.

\subsubsection{Attack Surface Analysis.}\label{sec:AttackSurface}
To capture a comprehensive attack surface for ML-enabled medical devices, we have developed tools for automated searching of public databases and identifying known vulnerabilities in the \textit{peripheral} and \textit{interconnected medical devices}. Comprehensive attack surface analysis is a prerequisite for systems-theoretic security analysis

In our recent work~\cite{Hallajiyan2024SAM}, we developed a method for capturing all the known vulnerabilities linked to each technology in every peripheral device used in a given ML-enabled medical device. This method uses the responses about the technological and operational factors used in the peripheral devices from the questionnaires (see \S\ref{sec:device-modeling}) as search keywords to find known vulnerabilities in the MITRE Common Vulnerability Enumeration (CVE) database~\cite{cvedata}. 
For instance, for the d-Nav BGMS~\cite{dnav}, we found that vulnerabilities might exist in a compatible glucose meter~\cite{garbelini2020sweyntooth}, Wi-Fi communication between glucose meter and the cloud server~\cite{CVE-2020-26145}, the communication between the cloud server for glucose meter and the ML controller~\cite{dlink-vul}, Wi-Fi communication between the interface device and the ML controller~\cite{CVE-2020-26145}, and Android OS on the interface~\cite{CVE-2024-43093}.


In another study~\cite{xu2019analysis}, we examined cyberattacks targeting hospital networks and interconnected clinical environments. For this purpose, we used two publicly available vulnerability databases -- the Common Vulnerabilities and Exposures (CVE) Database~\cite{cvedata} and the Industrial Control Systems Cyber Emergency Response Team (ICS-CERT) Alerts database~\cite{ICS-CERT}. To automate the collection of information on medical device-related vulnerabilities from ICS-CERT, we developed a tool for crawling the whole US-CERT website and extracting all vulnerability records that contain any medical-related keywords, including generic medical keywords and those describing the common categories and specialties of medical devices, as classified by the FDA Product Code Classification Database~\cite{FDACodes}. Using this tool, we extracted the vulnerability records reported from 1999 to 2018 that were potentially related to medical devices. We then manually parsed the HTML documents of a final set of 140 extracted records to extract information such as the corresponding CVE IDs, 
affected product names, and manufacturer or vendor names of products, as well as vulnerability details and backgrounds. Our analysis revealed that the most common vulnerabilities included improper credential management and authentication, weak access control, privilege escalation, and buffer and stack overflows. Furthermore, we found that 18 retrieved vulnerabilities had publicly available exploits. These vulnerabilities were widespread across various medical devices, including insulin pumps, from multiple manufacturers, thereby underscoring the need to consider them in the security analysis of interconnected ML-enabled devices.

\begin{table}[!ht]
\captionsetup{justification=centering}
\caption{Examples of recalls of AI/ML-enabled medical devices due to software issues that could result in misdiagnosis or wrong treatment (Class II recalls).\\{\footnotesize $^*$ indicates that the recalls are in \textit{Open} state as of April 2025, meaning that not all the units have been corrected or removed yet.}}
\label{tab:mlrecalls}
\scriptsize
\setlength\tabcolsep{1pt}
\centering
\begin{tabular}{|c|c|c|c|c|c|c|c|c|}
\hline
\rowcolor[HTML]{EFEFEF} 
  \textbf{\begin{tabular}[c]{@{}c@{}}Device\\Name\end{tabular}} &
  \textbf{\begin{tabular}[c]{@{}c@{}}Approval\\Panel\end{tabular}} &
  \textbf{\begin{tabular}[c]{@{}c@{}}Recall\\\#\end{tabular}} &
  \textbf{\begin{tabular}[c]{@{}c@{}}Reason for\\Recall\end{tabular}} &
  \textbf{\begin{tabular}[c]{@{}c@{}}Action\\Summary\end{tabular}} &
  \textbf{\begin{tabular}[c]{@{}c@{}}No. of\\units\\affected\end{tabular}} \\ \hline
  \begin{tabular}[c]{@{}c@{}}BodyGuardian\\Heart\\Remote\\Monitoring\\Kit
\end{tabular} &
  \begin{tabular}[c]{@{}c@{}}Cardio-\\vascular\end{tabular} &
  \begin{tabular}[c]{@{}c@{}}Z-\\2479-\\2020\\\cite{bodyguardianrecall}\end{tabular} &
  \begin{tabular}[c]{@{}l@{}}The device data\\being collected and\\transferred to the\\ monitoring center may\\not be accurate due\\to non-validated\\association between\\the phone software\\and the heart\\monitors, leading to\\inaccurate evaluation\\of the patients'\\condition.\end{tabular} &
   \begin{tabular}[c]{@{}l@{}}The recalling firm contacted\\all patients and physicians\\that had potentially\\impacted devices. Patients\\that agreed were sent new\\devices to replace the\\affected one to finish their\\study.\end{tabular} &
  8 \\\hline
  \begin{tabular}[c]{@{}c@{}}Dario\\BGMS\\\end{tabular} &
  \begin{tabular}[c]{@{}c@{}}Clinical\\chemistry\end{tabular} &
  \begin{tabular}[c]{@{}c@{}}Z-\\0260-\\2020\\ \cite{dariorecall}\end{tabular} &
  \begin{tabular}[c]{@{}l@{}}The Dario Android\\App v4.3.0-4.3.2 may\\experience duplicate\\logging of a blood\\glucose level reading.\end{tabular} &
   \begin{tabular}[c]{@{}l@{}}The firm released Android\\App v4.3.3. Users were\\informed about the issue via\\multiple push notifications\\and email, asking them to\\update to the new version.\end{tabular} &
  126,271 \\\hline
  \begin{tabular}[c]{@{}c@{}}Bioplex\\2200 ANA\\Screen\\\end{tabular} &
  \begin{tabular}[c]{@{}c@{}}Clinical\\toxicology\end{tabular} &
   \begin{tabular}[c]{@{}c@{}}Z-\\1159-\\2008\\ \cite{bioplexrecall}\end{tabular} &
   \begin{tabular}[c]{@{}l@{}}False negative results\\due to reagent packs\\ exhibiting low signal.\end{tabular} &
   \begin{tabular}[c]{@{}l@{}}The firm contacted its\\consignees, informing them\\ of the issue, recommended\\that they perform QC\\testing daily with each\\reagent pack, and updated\\the usage instructions.\end{tabular} &
  8,804\\\hline
  \begin{tabular}[c]{@{}c@{}}Sight\\OLO\\CBC Test\\Kit\end{tabular} &
  \begin{tabular}[c]{@{}c@{}}Hematology\end{tabular} &
   \begin{tabular}[c]{@{}c@{}}Z-\\2173-\\2024$^*$\\ \cite{sightrecall}\end{tabular} &
   \begin{tabular}[c]{@{}l@{}}The kit shows a bias\\in the platelet count\\due to bacterial\\ contamination, which\\can result in elevated\\counts with a bias,\\that results in the\\test kit performing\\outside of the device\\specification.\end{tabular} &
   \begin{tabular}[c]{@{}l@{}}The manufacturer issued an\\urgent recall notice to their\\customers, asking them to\\discontinue the use of\\the affected test kits, return\\the unused kits, and dispose \\of the used ones.\end{tabular} &
  	7,450 \\\hline
  \begin{tabular}[c]{@{}c@{}}UniCel\\DxH 600\\Coulter\\Cellular\\Analysis\\ System\end{tabular} &
  \begin{tabular}[c]{@{}c@{}}Micro-\\biology\end{tabular} &
   \begin{tabular}[c]{@{}c@{}}Z-\\2158-\\2017\\ \cite{unicelrecall}\end{tabular} &
   \begin{tabular}[c]{@{}l@{}}A possible data\\acquisition disruption\\may cause some\\ unusual events, that\\may be incorrectly\\removed from analysis,\\which can result in\\ erroneous diagnosis.\end{tabular} &
   \begin{tabular}[c]{@{}l@{}}The manufacturers sent an\\Urgent Medical Device Recall\\letter to customers to inform\\them of the issue, impact,\\action, and resolution.\end{tabular} &
  1,408 \\\hline
  \begin{tabular}[c]{@{}c@{}}Incisive\\CT,(728143,\\728144),\\Software\\v5.0.0.\\\end{tabular} &
  \begin{tabular}[c]{@{}c@{}}Radiology\end{tabular} &
   \begin{tabular}[c]{@{}c@{}}Z-\\0640-\\2024$^*$\\ \cite{incisiverecall}\end{tabular} &
   \begin{tabular}[c]{@{}l@{}}Multiple software\\issues have the\\potential to lead to\\ misdiagnosis due to\\image artifacts or\\incorrect image\\orientation labels, or\\ need for a CT rescan. 
\end{tabular} &
   \begin{tabular}[c]{@{}l@{}}The manufacturer\\communicated specific details\\regarding the issue to their\\customer, as well as advice on\\actions to be taken. They also\\promised to install a software\\upgrade.\end{tabular} &
  828 \\\hline 
\end{tabular}  
\setlength\tabcolsep{6pt}
\vspace{1em}
\end{table} 


\begin{table}[!ht]
\captionsetup{justification=centering}
\caption{Examples of adverse events of AI/ML-enabled medical devices with data, interface device, and software related problems that could result in misdiagnosis or wrong treatment (Event Types: Malfunction). \\{\footnotesize $^*$ indicates several similar adverse events reported for the same device over 2018-2024.}}
\label{tab:mlMAUDE}
\scriptsize
\setlength\tabcolsep{1.5pt}
\centering
\begin{tabular}{|c|c|c|c|c|l|}
\hline
\rowcolor[HTML]{EFEFEF} 
\textbf{\begin{tabular}[c]{@{}c@{}}Device\\ Name\end{tabular}} & 
\textbf{\begin{tabular}[c]{@{}c@{}}Device\\ Function\end{tabular}} & 
\textbf{\begin{tabular}[c]{@{}c@{}}Approval\\ Panel\end{tabular}} & 
\textbf{\begin{tabular}[c]{@{}c@{}}Adverse\\ Event \#\\ (Year)\end{tabular}} & 
\textbf{\begin{tabular}[c]{@{}c@{}}Device \\ Problem\end{tabular}} & 
\multicolumn{1}{c|}{\cellcolor[HTML]{EFEFEF}\textbf{\begin{tabular}[c]{@{}c@{}}Summary \\ Event Description\end{tabular}}} \\ \hline

\begin{tabular}[c]{@{}c@{}}Zio AT \\ ECG Monitoring\\ System (ZEUS)\end{tabular} & \begin{tabular}[c]{@{}c@{}}Arrhythmia \\ detector\\ and alarm\end{tabular} & \begin{tabular}[c]{@{}c@{}}Cardio-\\ vascular\end{tabular} & \begin{tabular}[c]{@{}c@{}}8356453\\ \cite{MDR_8356453}\\ (2019)\end{tabular} & \begin{tabular}[c]{@{}c@{}}Application \\ Network\\ Problem\end{tabular} & \begin{tabular}[c]{@{}l@{}}False negative results \\ (missed detection of \\ asymptomatic arrhythmia) \\ due to a BLE (bluethooth \\ low energy) issue.\end{tabular} \\ \hline

\begin{tabular}[c]{@{}c@{}}LINQ II \\ Cardiac Monitor, \\ Zelda AI ECG \\ Classification \\ System\end{tabular} & 
\begin{tabular}[c]{@{}c@{}}Arrhythmia \\ detector\\ and alarm\end{tabular} & 
\begin{tabular}[c]{@{}c@{}}Cardio-\\ vascular\end{tabular} & 
\begin{tabular}[c]{@{}c@{}}20916084\\ \cite{MDR_20916084}\\ (2024)\end{tabular} & 
\begin{tabular}[c]{@{}c@{}}Program or \\ Algorithm \\ Execution\\ Problem\end{tabular} & 
\begin{tabular}[c]{@{}l@{}}An atrial fibrillation \\ episode was adjudicated \\ as false by the artificial \\ intelligence (ai) algorithm. \end{tabular} \\ \hline

\begin{tabular}[c]{@{}c@{}}Dario\\BGMS\\ \end{tabular} & 
\begin{tabular}[c]{@{}c@{}}Glucose \\Test \\System \end{tabular} & 
\begin{tabular}[c]{@{}c@{}}Clinical\\chemistry \end{tabular} & 
\begin{tabular}[c]{@{}c@{}}18904273\\ \cite{MDR_18904273}\\ (2022)$^*$\end{tabular} & 
\begin{tabular}[c]{@{}c@{}} Incorrect, \\ Inadequate, \\ or Imprecise \\ Result or \\ High Readings\end{tabular} & 
\begin{tabular}[c]{@{}l@{}} Inconsistent and high \\ blood glucose readings, \\ different from other meters \\ or hospital measurements. \end{tabular} \\ \hline

\begin{tabular}[c]{@{}c@{}}HeartFlow\\FFRCT\\ \end{tabular} & 
\begin{tabular}[c]{@{}c@{}}Coronary \\ Vascular \\ Physiologic \\ Simulation \\ Software\end{tabular} & 
\begin{tabular}[c]{@{}c@{}}Cardio-\\ vascular \end{tabular} & 
\begin{tabular}[c]{@{}c@{}}8269286\\ \cite{MDR_8269286}\\ (2018)$^*$\end{tabular} & 
\begin{tabular}[c]{@{}c@{}} False \\ Negative \\ Result \end{tabular} & 
\begin{tabular}[c]{@{}l@{}} Potential false negative \\ results in FFRCT \\ (Fractional Flow Reserve \\ derived from CT) analysis \\ of coronary arteries due to \\ image quality issues and \\ anatomy uncertainty. \end{tabular} \\ \hline

\begin{tabular}[c]{@{}c@{}} Clarius \\ Ultrasound \\ Scanner \end{tabular} & 
\begin{tabular}[c]{@{}c@{}} Ultrasonic \\pulsed \\ doppler \\imaging \\ system\end{tabular} & 
\begin{tabular}[c]{@{}c@{}}Radiology \end{tabular} & 
\begin{tabular}[c]{@{}c@{}}\\20471171 \\ \cite{MDR_20471171}\\ (2024)\end{tabular} & 
\begin{tabular}[c]{@{}c@{}} Misconnection \end{tabular} & 
\begin{tabular}[c]{@{}l@{}} A connectivity issue with \\ ultrasound scanner during \\ diagnostic evaluation in \\ an emergency room, which \\ could potentially lead to \\ significant adverse outcomes.  \end{tabular} \\ \hline
\end{tabular}%
\end{table}
\subsubsection{Analysis of Recalls and Adverse Events in Medical Devices.} In our early work~\cite{Alemzadeh_SP,alemzadeh2014automated,alemzadeh2016adverse}, we developed a suite of NLP tools (called MedSafe~\cite{MedSafeRecall,MedSafeMAUDE}) for automated extraction, cross-referencing, and classification of records from two public FDA databases: the Medical Device Recalls~\cite{FDA_Recalls} and the MAUDE (Adverse Event Reports) database~\cite{FDA_MAUDE}. We used these tools to identify all the recalls and adverse events caused by failures in computer-based medical devices, and categorized them by fault class, failure mode, device type, recovery action, and the number of recalled devices. This study was the first automated and large-scale analysis of FDA data on computer-based medical devices and highlighted the key causes of computer failures impacting patient safety.
Our findings showed that while software failures continue to be the leading cause of medical device failures, hardware, battery, and I/O issues are also major contributors. Many recalled devices either lacked proper safety considerations during design or their safety mechanisms were inadequately implemented. Later, using these tools we extracted all the recalls and adverse events related to BGMS~\cite{zhou2022design,zhou2023hybrid} and surgical robots~\cite{alemzadeh2016adverse} and identified most common device malfunctions, examples of security vulnerabilities in different components and device interfaces (e.g., CGMs, insulin pumps, cameras), and the safety impact of device failures and potential harm to patients (e.g., hyperglycemia or injury). For example, we found a Class 1 recall (highest risk level) due to a potential security vulnerability related to the use of the remote controller accessories with the insulin pumps, which affected over 90,000 users on the market~\cite{MMT500recall}. Another Class 2 recall, affecting over 64,000 insulin pumps, indicated the possibility of an unauthorized person connecting wirelessly to a nearby insulin pump to change settings and control insulin delivery due to potential cybersecurity vulnerabilities~\cite{MiniMedRecall}.

\begin{figure}[t!]
\begin{minipage}[b]{0.48\linewidth}
    \centering
   \includegraphics[width=\linewidth]{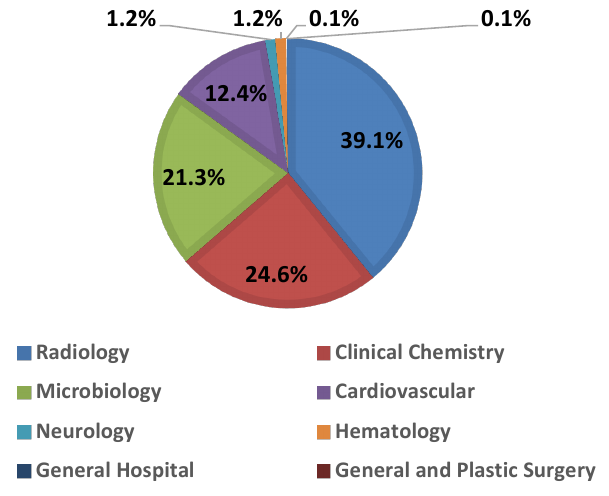} 
\end{minipage}%
\hfill
\begin{minipage}[b]{0.42\linewidth}
    \scriptsize
    \setlength\tabcolsep{4pt}
    \centering
    \begin{tabular}{|l|c|}
    \hline
    \rowcolor[HTML]{EFEFEF} 
    \textbf{\begin{tabular}[c]{@{}c@{}}Reported Device Problem\end{tabular}} & 
    \textbf{\begin{tabular}[c]{@{}c@{}}Freq.\end{tabular}} \\
    \hline
    {\begin{tabular}[l]{@{}l@{}}Poor Quality Image\end{tabular}} & 
    {\begin{tabular}[l]{@{}l@{}}421\end{tabular}} \\ \hline
    {\begin{tabular}[l]{@{}l@{}}Not Available (N/A)\end{tabular}} & 
    {\begin{tabular}[l]{@{}l@{}}295\end{tabular}} \\ \hline
    {\begin{tabular}[l]{@{}l@{}}False Negative Result\end{tabular}} & 
    {\begin{tabular}[l]{@{}l@{}}103\end{tabular}} \\  \hline
    {\begin{tabular}[l]{@{}l@{}}High Readings\end{tabular}} & 
    {\begin{tabular}[l]{@{}l@{}}73\end{tabular}} \\ \hline
    {\begin{tabular}[l]{@{}l@{}}Incorrect, Inadequate, or \\ Imprecise Result or Readings\end{tabular}} & 
    {\begin{tabular}[l]{@{}l@{}}45\end{tabular}} \\  \hline
    {\begin{tabular}[l]{@{}l@{}}Computer Software Problem\end{tabular}} & 
    {\begin{tabular}[l]{@{}l@{}}42\end{tabular}} \\  \hline
    {\begin{tabular}[l]{@{}l@{}}Inaccurate Information\end{tabular}} & 
    {\begin{tabular}[l]{@{}l@{}}25\end{tabular}} \\  \hline
    {\begin{tabular}[l]{@{}l@{}}Application Program Problem: \\ Parameter Calculation Error\end{tabular}} & 
    {\begin{tabular}[l]{@{}l@{}}17\end{tabular}} \\  \hline
    {\begin{tabular}[l]{@{}l@{}}Low Readings\end{tabular}} & 
    {\begin{tabular}[l]{@{}l@{}}7\end{tabular}} \\  \hline
    {\begin{tabular}[l]{@{}l@{}}Intermittent Program or\\ Algorithm Execution\end{tabular}} & 
    {\begin{tabular}[l]{@{}l@{}}6\end{tabular}} \\  \hline
    {\begin{tabular}[l]{@{}l@{}}Program or Algorithm\\ Execution Failure\end{tabular}} & 
    {\begin{tabular}[l]{@{}l@{}}4\end{tabular}} \\  \hline
    {\begin{tabular}[l]{@{}l@{}}Failure to Transmit Record\end{tabular}} & 
    {\begin{tabular}[l]{@{}l@{}}3\end{tabular}} \\  \hline
    {\begin{tabular}[l]{@{}l@{}}Low Test Results\end{tabular}} & 
    {\begin{tabular}[l]{@{}l@{}}3\end{tabular}} \\  \hline
    \end{tabular}
    \label{tab:top_MDRs}
\end{minipage}
\captionsetup{justification=centering}
\caption{Left: Adverse Events by Device Category (FDA Approval Panel), \\Right: Top Reported ML-enabled Device Problems (Data as of April 2025)}
    \label{fig:MDR_Categories}
\end{figure}

More recently, we have applied our techniques to extract and analyze the recalls and adverse events reported on ML-enabled devices and AID systems. Our analysis found over 1,460 adverse events reported for ML-enabled devices over 2015-2024, of which about 92\% involved device malfunctions and 7.8\% injuries. Although understanding the root causes of the reported events requires an in-depth investigation and consideration of all causal and contextual factors~\ref{sec:CAST}, these reports provide valuable insights on real problems encountered during the use of devices and how they impacted patient safety. Figure \ref{fig:MDR_Categories} shows the device categories with the highest number of adverse event reports and the top device problems reported over the years. A major part of reported problems (about 60.7\%) were related to poor quality and inaccurate inputs/readings and false negative results, which, even if not directly caused by an ML technology, could still impact the ML decision-making results and patient safety. Some examples of safety-critical recalls and adverse events across different device categories are shown in Tables \ref{tab:mlrecalls} and \ref{tab:mlMAUDE}.


In summary, this set of tools and techniques would help a security analyst cover known vulnerabilities in ML models, peripheral medical devices, as well as attack vectors in connected peripheral devices and communication channels, while designing a \textit{secure} ML-enabled medical device. Additionally, it will also help security practitioners design efficient attack prevention and detection techniques.

\subsection{Data-Driven Systems-Theoretic Safety and Security Analysis} 
To predict and proactively mitigate the occurrence of future attacks, it is crucial to not only consider the known vulnerabilities and exploits reported in existing data on past safety and security incidents, but also anticipate for the potential new attacks by considering a more comprehensive attack surface of unknown vulnerabilities or vulnerabilities in other connected devices and the potential attack steps and their safety impacts on patients. To do this, we adopt an LLM-aided and data-driven approach to the systems-theoretic security analysis (STPA-Sec) that incorporates the information extracted from public databases and results from CAST analysis on devices with similar functionality (e.g., a non-ML-enabled device predicate with the same functional specification and use cases as the ML-enabled device) to generate potential attacks steps and their impacts in ML-enabled devices. 

\subsubsection{Systems-Theoretic Accident Causality (CAST) Analysis.}\label{sec:CAST}
Analysis of real-world safety incidents, including medical device recalls~\cite{FDA_Recalls} and adverse events~\cite{FDA_MAUDE} can provide valuable insights into how device flaws and security vulnerabilities could lead to system hazards and negatively impact patients and caregivers. However, these incidents are mainly reported by the device users and manufacturers in free-form natural language text, and their analysis requires a semantic understanding of the underlying causal factors. Several previous studies~\cite{balgos2012systems,alemzadeh2013safety,o2014application,mason2017assessing,canham2018examining,leveson2020systems} have shown the advantage of Causal Analysis using System Theory (CAST)~\cite{leveson2011engineering} in identifying causal and contextual factors contributing to medical adverse events. However, these papers solely focus on the manual causal analysis of single incidents and do not consider security-related hazards and safety-critical vulnerabilities. Such an approach cannot provide a comprehensive understanding of all potential causal factors, including vulnerabilities in IoT and peripheral devices, nor can it yield statistically significant measures of their importance. Additionally, it is not easily scalable to thousands of unstructured adverse event reports on a single device due to the significant human effort required. Therefore, techniques and tools for automated semantic analysis of these reports are needed to extract both safety and security-related causal factors, and summarize key information for CAST analysis.

\begin{figure}[t!]
    \centering
     \includegraphics[width=\columnwidth,keepaspectratio]{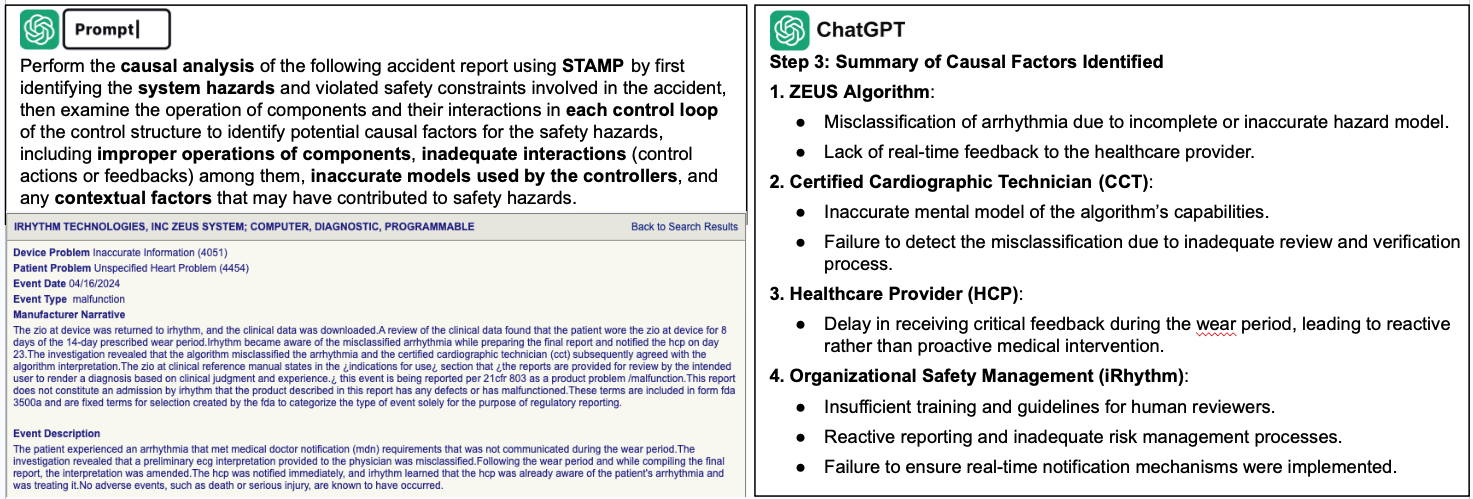}
    \captionsetup{justification=centering}
    \caption{Example Accident Causality Analysis using STAMP for an FDA-Authorized ML-enabled Medical Device}
    \label{fig:CAST_example}
    \vspace{-1em}
\end{figure}

To facilitate an aggregated CAST analysis, we leverage our NLP techniques for automated classification, summarization, and cross-referencing of large-scale FDA data on recalls and adverse events~\cite{Alemzadeh_SP,alemzadeh2014automated,alemzadeh2016adverse,zhou2022design,elnawawy2024systematically}. This information can assist in systems-theoretic analysis of several similar adverse events reported on the same medical device or devices with the same functional specification using CAST to identify the distribution of causal factors and potentially inadequate safety mechanisms in both system design and operational practices~\cite{alemzadeh2016thesis,alemzadeh2015systems}. Given a natural language description of an adverse event and the control structure model for a medical device, we first map different sections of the text into different control loops in the system control structure. Then, for each control loop, the set of violated safety constraints is identified. These steps are done through device and medical entity and relation extraction from the text and semantic analysis of causal factors using rule-based parts of speech analysis~\cite{alemzadeh2016thesis}. Finally, the similar causal factors and safety violations across multiple adverse event reports of the same device can be identified and aggregated for statistical analysis. In ~\cite{alemzadeh2016adverse,alemzadeh2015systems} we performed such an aggregated causality analysis of over 10,000 adverse event reports on tele-operated surgical robots. This analysis identified the most critical causal factors for safety incidents in different models of the same device over a period of 14 years. To further reduce the manual cost of this analysis method, we have recently explored an LLM-aided technique based on customized prompt templates and chain-of-thought prompting~\cite{ge2023dkec,wei2022chain} to decompose the tasks of entity and relation extraction and semantic analysis of causal factors into sub-tasks that can be performed using LLMs and be later manually validated by security analysts. Figure~\ref{fig:CAST_example} shows an example of the key causal factors extracted by this LLM-aided technique from an adverse event report~\cite{FDA_ZEUS_MAUDE} for an FDA-authorized ML-enabled cardiac event detection software~\cite{FDA_ZEUS}.
The insights on the causes and patient impacts of past incidents can be used for analyzing and specifying the safety impact of the device vulnerabilities.

\subsubsection{Systems-Theoretic Security (STPA-Sec) Analysis.} \label{sec:stpa-sec}
In this final step, we analyze consolidated data on the ML model, its functionality, peripheral technologies, and associated safety and security risks to identify how an adversary could inject false data during inference. We developed STPA-Sec for ML-enabled Medical Devices (\stpatool{}), a technique for conducting STPA-Sec on AI/ML-enabled medical devices~\cite{Hallajiyan2024SAM}. \stpatool{} first assesses the attack surface by identifying all potential attack entry points (Section \ref{sec:AttackSurface}). 
Thereafter, it performs STPA-Sec analysis to determine the attack steps. This information would help the ML-enabled device manufacturer design appropriate security measures or devise advisories for the users.

In the attack step generation step, \stpatool{} performs an LLM-aided STPA-Sec analysis to generate the attack steps for a given hazard and an exploitable peripheral device vulnerability. 
To overcome a human security analyst's limited cross-domain knowledge, we leverage LLMs to automatically identify causal scenarios based on the latest vulnerabilities in the system's underlying technologies. 
A key challenge in using LLMs is the design of effective prompts to generate optimal task-specific responses. For \stpatool{}, the ideal response outlines detailed attack steps exploiting a peripheral vulnerability to inject false data during inference on a given ML technique. To achieve this, we developed the following prompt. 
\\
\begin{tcolorbox}[breakable,colback=light-gray,arc=0pt,outer arc=0pt,colframe=light-gray]
``Act as a security engineer who has the task of identifying the steps that an adversary follows to cause a security breach in an ML-enabled medical system. $<$\textit{Description of an ML-enabled medical system}$>$. $<$\textit{Definition of security breach}$>$. You are given a system description, an ML attack, a targeted input peripheral component, and a known vulnerability in the input component. Give a list of steps to show how an adversary can exploit the vulnerability to mislead the ML-enabled component and how that affects the action of the output device on the patient.

\noindent\textbf{System Description:} $<$\textit{The \stpatool{} user manually writes this description by inspecting information disclosed by the manufacturer.}$>$

\noindent\textbf{Data flow:} $<$\textit{This can be derived from the control structure constructed using the Control structure builder in \S\ref{sec:device-modeling}}.$>$

\noindent\textbf{ML attack:} $<$\textit{The ML attack identified in \S\ref{sec:infoextraction}}$>$

\noindent\textbf{Targeted input peripheral component:} $<$\textit{One of the peripheral input devices in the control structure built in \S\ref{sec:device-modeling}}$>$

\noindent\textbf{Targeted technology:} $<$\textit{One of the underlying technologies in the input device, as identified by the technology identifier (\S\ref{sec:device-modeling})}$>$

\noindent\textbf{Known vulnerability:} $<$\textit{Description of the known vulnerability in the targeted technology, as retrieved from the CVE database during attack surface analysis}$>$''\\
\end{tcolorbox}

We observed that explicitly assigning the LLM the role of a security analyst before giving it additional information improves the readability and relevance of the generated results - this is in line with other work in this area~\cite{shanahan2023role,li2024camel,santu2023teler}. Similarly, mentioning the data flow provides clarity to the LLM regarding the sequence of data transmission between different components in the system.  

By running this prompt for each vulnerable point in the system and each vulnerability uncovered at that point, \stpatool{}, regardless of the existence of safety/security margins, generates a comprehensive set of steps an adversary might take to compromise the security of an ML-enabled medical device. Device manufacturers or security analysts can then disregard those that have already been mitigated and develop design recommendations for the remaining ones.


For d-Nav, we selected hypoglycemia as the hazard, and \textit{injecting excess insulin} as the control action that causes it. We consider an adversary who conducts a model inversion attack (identified by MedAIScout~\cite{dharmalingammedaiscout} in a previous step described in \S\ref{sec:infoextraction}) on the ML engine to infer sensitive details about a targeted patient, followed by false data injection. This attack would make the ML engine mispredict the insulin dose. To execute this attack, the adversary injects false glucose readings into the Wi-Fi channel that transmits the patient's glucose readings from the glucose meter to the ML engine running in the cloud server. We assume that the patient uses a Wi-Fi router with an unpatched known vulnerability, \textit{CVE-2023-35836} \cite{dnavcve}, that the adversary exploits for injecting the malicious glucose readings. 
SAM outputs a list of nine steps for this attack, which are summarized in Table~\ref{tab:dnavattack}. By following these steps, an adversary could inject false data into the BGMS to make it miscalculate the insulin dose. 

\begin{table}[h]
\captionsetup{justification=centering}
\caption{STPA-Sec output produced by SAM for the attack scenario on d-Nav BGMS, described in \$\ref{sec:stpa-sec}}
    \label{tab:dnavattack}
    \centering
    \begin{tabular}{|c|c|c|}
        \hline
        \rowcolor[gray]{0.8} 
        \textbf{Step \#} & \textbf{Step name} & \textbf{Description} \\
        \hline
        1 & Reconnaissance & \begin{tabular}[c]{@{}c@{}}Identifying the targeted patient's Wi-fi\\network and its router vulnerabilities\end{tabular} \\
        \hline
        2 & Exploitation & \begin{tabular}[c]{@{}c@{}}Exploitation of router vulnerability to\\infiltrate the target's network\end{tabular} \\
        \hline
        3 & Wi-fi network infiltration & \begin{tabular}[c]{@{}c@{}}Compromising the connection between\\glucose meter and cloud server\end{tabular} \\
        \hline
        4 & Data interception & Interpreting the data in transit\\
        \hline
        5 & Data tampering & \begin{tabular}[c]{@{}c@{}}Manipulating the data in transit with a\\value that would make the ML model\\mispredict a future blood glucose value\end{tabular} \\
        \hline
        6 & Model inversion attack & \begin{tabular}[c]{@{}c@{}}Compute the manipulated value such\\that the patient becomes hypoglycemic\end{tabular}\\
        \hline
        7 & ML Controller manipulation & \begin{tabular}[c]{@{}c@{}}Expected reaction of the ML model:\\ Misprediction of patient's future blood\\glucose level\end{tabular}\\
        \hline
        8 & Output device manipulation & \begin{tabular}[c]{@{}c@{}}Expected reaction of the insulin dose\\ calculator: Computing an insulin dose\\higher than that required by the patient\\and sent to the insulin pump or displayed\\on the d-Nav app\end{tabular} \\
        \hline
        9 & Insulin pump misadministration & \begin{tabular}[c]{@{}c@{}}Expected end result: Wrong insulin dose\\ administered to the patient, either\\manually or by an automated insulin pump\end{tabular} \\
        \hline
    \end{tabular}
\end{table}



\section{Future Directions and Conclusion}\label{sec:future-directions}
Based on the capabilities of our tools and techniques demonstrated in this paper, and the insights obtained from the experimental results, we would like to expand the scope of our research in the following directions, with a high-level goal of ensuring the security of ML-enabled medical devices by design and efficient post-market security surveillance.
\begin{enumerate}[leftmargin=*]
    \item Early prediction of vulnerabilities based on existing events - The domain of ML-enabled medical devices has become increasingly competitive, with manufacturers developing ML-enabled devices that offer similar core functionalities as existing non-ML-enabled devices, but with enhanced performance and features such as greater interoperability. As a result, newer devices may inherit existing vulnerabilities in connected devices or similar or more severe vulnerabilities than their predecessor devices. To address this, we plan to develop an LLM-aided technique that analyzes the design of an ML-enabled medical device and, based on known vulnerability data, predicts potential security risks specific to the new device, even without performing STPA-Sec on it. Furthermore, we would expand the scope of our tools and techniques to cover other types of ML-specific attacks in addition to false data injection attacks.
    \item Real-time post-market security risk assessment - Our tools and techniques can be extended to support near real-time post-market security surveillance by continuously monitoring large-scale vulnerability databases and performing on-demand risk assessments whenever new vulnerabilities are reported.
    \item Designing efficient defense techniques - The output of our STPA-Sec technique can be leveraged to identify the most efficient defense technique in terms of reliability, patient convenience, and cost of implementation.
\end{enumerate}

This paper presents a suite of tools and techniques developed for holistic security risk assessment of ML-enabled medical devices, with a focus on false data injection attacks. We demonstrated the effectiveness of these tools and techniques across multiple ML-enabled blood glucose management systems. The novelty of our tools and techniques are in (i) identifying attack vectors that require exploiting vulnerabilities in third-party connected components to practically execute known attacks on the ML models and (ii) anticipating for the potential safety impacts of such attacks based on the analysis of past incidents on similar devices. This helps security analysts (working for the device manufacturers) assess the feasibility and impact of such attacks more accurately. In the future, we aim to extend our tools to support additional types of ML-specific attacks and facilitate post-market security risk assessments.

\begin{credits}
\subsubsection{\ackname} This research was partially supported by the Natural Sciences and Engineering Research Council of Canada (NSERC), the National Research Council of Canada’s (NRC) Digital Health and Geospatial Analytics Program, and the U.S. National Science Foundation (CNS-2146295). 

\subsubsection{\discintname}
The authors have no competing interests to declare that are
relevant to the content of this article. 
\end{credits}
%
%
%
\bibliographystyle{splncs04}
\bibliography{main.bib}

\end{document}